\documentclass[aps,prl,amsmath,twocolumn,showpacs,superscriptaddress]{revtex4-1}
\usepackage{amsmath, amssymb, graphicx, bm}
\usepackage[usenames,dvipsnames,svgnames,table]{xcolor}
\usepackage[colorlinks=true,linkcolor=RoyalBlue,citecolor=RoyalBlue]{hyperref}

\begin{document}
\title{Interlayer Couplings Mediated by Antiferromagnetic Magnons}

\author{Ran Cheng}
\affiliation{Department of Physics, Carnegie Mellon University, Pittsburgh, PA 15213}
\affiliation{Department of Electrical and Computer Engineering, Carnegie Mellon University, Pittsburgh, PA 15213}
\affiliation{Department of Electrical and Computer Engineering, University of California, Riverside, CA 92521}

\author{Di Xiao} 
\affiliation{Department of Physics, Carnegie Mellon University, Pittsburgh, PA 15213}

\author{Jian-Gang Zhu}
\affiliation{Department of Electrical and Computer Engineering, Carnegie Mellon University, Pittsburgh, PA 15213}
\affiliation{Department of Physics, Carnegie Mellon University, Pittsburgh, PA 15213}

\begin{abstract}

Collinear antiferromagnets (AFs) support two degenerate magnon excitations carrying opposite spin polarizations, by which magnons can function as electrons in various spin-related phenomena. In an insulating ferromagnet(F)/AF/F trilayer, we explore the magnon-mediated interlayer coupling by calculating the magnon thermal energy in the AF as a function of the orientations of the Fs. The effect manifests as an interlayer exchange interaction and a perpendicular magnetic anisotropy; they both depend on temperature and the AF thickness. In particular, the exchange interaction turns out to be antiferromagnetic at low temperatures and ferromagnetic at high temperatures, whose magnitude can be $10-100$ $\mu$eV for nanoscale separations, allowing experimental verification.

\end{abstract}

\maketitle

An emerging thrust in the broad field of magnonics~\cite{ref:Magnonics} has been the investigation of collinear antiferromagnets (AFs) as promising candidate materials. This is ascribed to an appealing feature unique to collinear AFs: magnon excitations always come up with two degenerate modes as a consequence of symmetry~\cite{ref:AFMR}. The two modes carry opposite spin polarizations and form an intrinsic degree of freedom, which enables antiferromagnetic magnons to function as active information carriers in replacement of electrons~\cite{ref:FET,ref:Chirality,ref:Matthew}. For example, a longitudinal temperature gradient can drive magnons with opposite spins to opposing transverse edges in layered transition metal trichalcogenides~\cite{ref:SNE1,ref:SNE2,ref:SNEexp}, realizing a magnonic analogue of the spin Hall effect. Very recently, we have proposed that antiferromagnetic magnons can be utilized to induce magnetic switching in a ferromagnet (F)/AF/F heterostructure~\cite{ref:magtorque}, paving the way toward all-magnonic memory devices. These studies have demonstrated the exciting possibility of exploiting antiferromagnetic magnons to fulfill tasks that used to be exclusive to electrons.

In spite of tremendous recent progress, the knowledge we have acquired so far refers largely to the transport property, namely, magnon spin currents generated by external stimuli~\cite{ref:Kai,ref:Rezende}. As for the static property, on the other hand, it remains unexplored whether the analogy between antiferromagnetic magnons and electrons still holds. In particular, when two Fs are separated by an \textit{insulating} thin-film AF, will magnons in the AF mediate nonlocal interactions between the Fs? Here, magnons are supposed to play the role of conduction electrons in the nonmagnetic metal (NM) layer of a F/NM/F sandwich~\cite{ref:Grunberg,ref:Parkin,ref:Edwards,ref:Bruno}. This problem becomes especially important if at least one of the Fs is also insulating so that electron tunneling~\cite{ref:tunnelRKKY1,ref:tunnelRKKY2} across the AF is absent.

To address the above issue, we study in this Letter a F/AF/F trilayer structure with exchange-coupled interfaces. We extract the effective interlayer coupling by computing the magnon thermal energy inside the AF as a function of the relative orientation of the two Fs at different AF thicknesses and temperatures. To avoid ambiguity, we only consider insulating magnets so that conduction electrons are eliminated. We allow the AF to have multidomains in the lateral dimensions but domain boundaries do not exist in the thickness direction~\cite{ref:magtorque}, which is true for epitaxially grown films. Although the F/AF interface is exchange coupled, we exclude the exchange bias (EB) effect in this Letter because our predictions can be viewed as additional effects on top of the EB. Moreover, it is experimentally possible to suppress the EB to a negligible level with proper materials choice and careful control of film growth~\cite{private}.

Let the thickness direction be $z$ and the transverse dimensions be labeled by $x$ and $y$. Within each domain, translational symmetry is respected along $x$ and $y$ so the in-plane wave vector $\bm{k}_\parallel=\{k_x,k_y\}$ is a good quantum number. To better associate the phenomenology to the antiferromagnetic character, we only consider a $G$-type AF with simple cubic lattice and compensated interfaces. One can easily show that an $A$-type AF or a complicated bipartite lattice can only introduce quantitative differences, leaving the essential physics unaltered.

\textit{Model.}---We assume that the Curie temperature of the Fs far exceeds the N\'{e}el temperature of the AF, which is typically the case for insulating Fs and AFs. As a result, when working below the N\'{e}el temperature, we ignore magnon excitations in the Fs and focus only on magnons in the AF, treating the two Fs as boundary conditions practically independent of temperature. For a specified domain with collinear ground state, the AF is characterized by the spin Hamiltonian
\begin{align}
 H_{\rm AF}=J_{\rm AF}\sum_{\langle ij\rangle}\bm{s}_i\cdot\bm{s}_j+K_{\rm AF}\sum_i(\bm{\hat{e}}\cdot\bm{s}_i)^2,
\end{align}
where $\bm{s}_i$ is the local spin vector on site $i$, $J_{\rm AF}>0$ is the nearest-neighbor Heisenberg exchange interaction, $K_{\rm AF}<0$ is the anisotropy constant, $\bm{\hat{e}}$ is the local easy axis (uniform inside a given domain), and $\langle ij\rangle$ denotes nearest neighbor links. On the interfaces, the coupling between F and AF is represented by
\begin{align}
 H_{\rm int}=J_{\rm int}\sum_{i; I}\bm{s}_i\cdot\bm{S}_I,
\end{align}
where $\bm{S}_I$ ($I=1,2$) are the unit magnetization vectors of the Fs and $J_{\rm int}$ is the interfacial exchange coupling connecting the AF to the adjacent F. In the following, we normalize energies and temperature $k_BT$ to $J_{\rm AF}$, which is the largest energy scale in the system. In particular, we set $K_{\rm AF}/J_{\rm AF}=-0.05$~\cite{noteratio} and let $\xi=J_{\rm int}/J_{\rm AF}$ be the dimensionless interfacial coupling ($\xi$ is tunable). The sign of $\xi$ does not matter because the magnon-mediated interactions turn out to be scaled as $\xi^2$.

\begin{figure}[t]
	\includegraphics[width=0.95\columnwidth]{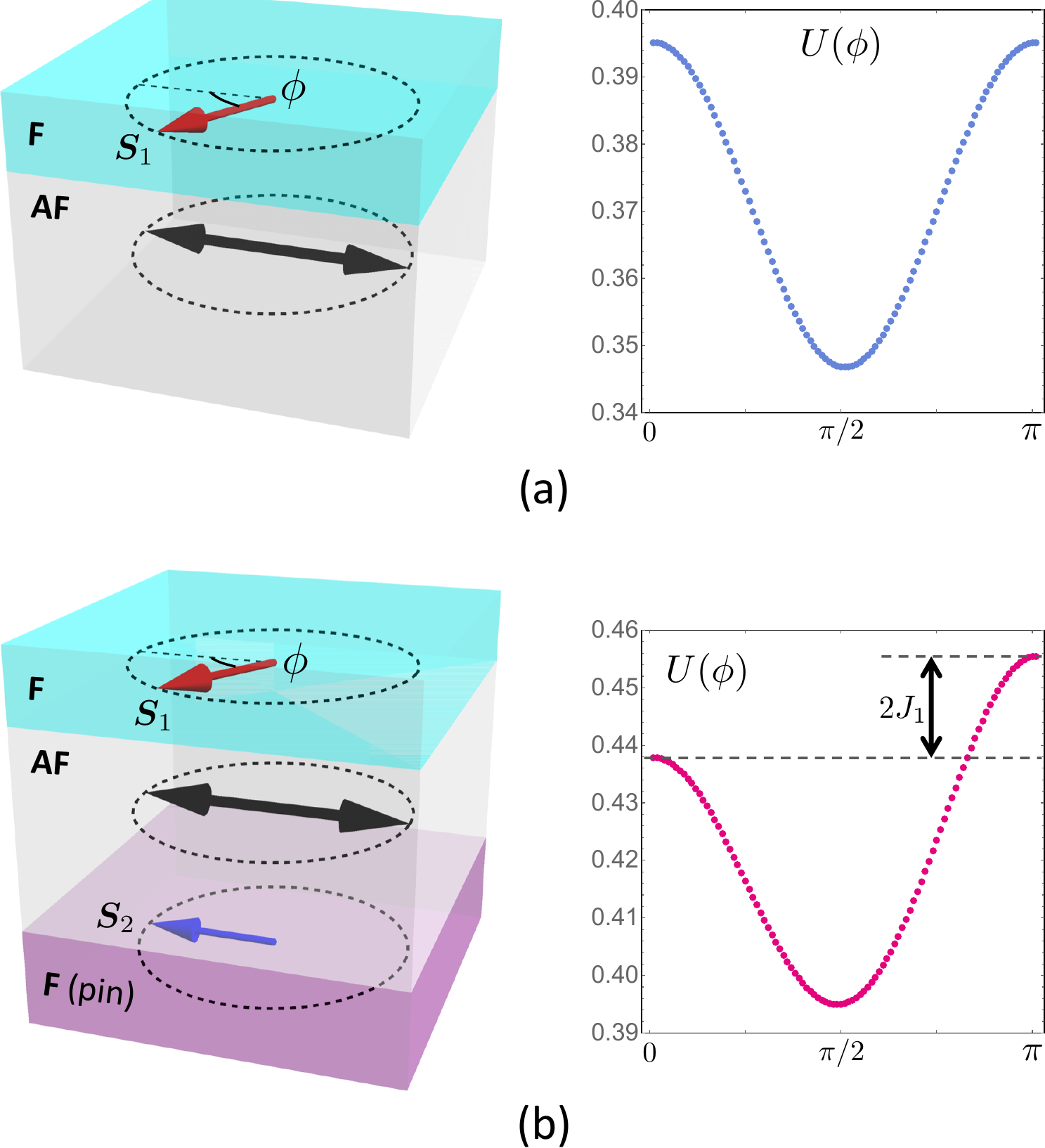}
	\caption{Dependence of the internal energy $U$ on the free layer orientation in: (a) F/AF bilayer; (b) F/AF/F trilayer with the bottom F pinned parallel to the N\'{e}el vector.}
	\label{fig:layers}
\end{figure}

Using the Holstein-Primakoff transformation~\cite{ref:Nolting}, we can recast the total Hamiltonian $H=H_{\rm AF}+H_{\rm int}$ in a tight-binding model~\cite{ref:SNE1} that parametrically depends on $\bm{S}_1$ and $\bm{S}_2$. Then magnons are similar to particles confined in a quantum well. For a given configuration of $\bm{S}_1$ and $\bm{S}_2$, \textit{i.e.}, boundary conditions, the magnon spectrum is solved as $\varepsilon_i(\bm{k}_\parallel)$ with discretized band index $i$ and continuous $\bm{k}_\parallel$. The internal energy of the AF consists of thermal excitations of all magnon bands obeying the Bose-Einstein distribution. Consequently, $U=U(\bm{S}_1,\bm{S}_2)$ is also a function of $\bm{S}_1$ and $\bm{S}_2$, which tells the effective interactions between the two Fs. However, bilayer effects associated with each individual F/AF interface also contribute to $U$. In order to separate the desired interlayer coupling (a trilayer effect) from $U$, we must compare a F/AF bilayer with a F/AF/F trilayer as illustrated in Fig.~\ref{fig:layers}. The difference between the two cases can be attributed to the effective interlayer coupling after an overall energy shift is disregarded.

\textit{Results.}---Let us consider a simple circumstance where the AF is single domain with an in-plane N\'{e}el vector, and the free F ($\bm{S}_1$) is restricted to rotate in the plane. In the trilayer case, a bottom F ($\bm{S}_2$) is added and fixed parallel to the N\'{e}el vector. So, in both cases [Figs.~\ref{fig:layers} (a) and~(b)] the internal energy $U$ has a single argument $\phi$. Thermodynamically, the amount of change in $U(\phi)$ by varying $\phi$ is regarded as the work done by the free F on the AF. By symmetry, $U(\phi)$ can be expanded as
\begin{align}
 \mbox{Bilayer: }  &U_B(\phi)=U_0+\sum_{n=1}^{\infty}\mathcal{K}_n\cos2n\phi, \label{eq:Ks} \\
 \mbox{Trilayer: } &U_T(\phi)=U_0'-\sum_{n=1}^{\infty}J_n\cos n\phi, \label{eq:Js}
\end{align}
where $U_0$ and $U_0'$ are two reference energies that do not concern us here. In Eq. (4), we adopted the minus-sign convention such that $J_n>0$ ($<0$) refers to ferromagnetic (antiferromagnetic) interactions. In $U_B$, we interpret $\mathcal{K}_n$ as a series of magnon-induced anisotropies. The $\phi$-dependence of the energy difference $\Delta U=U_T-U_B$ is characterized by $J_n$ ($n\in$odd) and $\tilde{J}_n=J_n+\mathcal{K}_{n/2}$ ($n\in$even), which we interpret as the magnon-mediated interlayer couplings. To fit those parameters, we numerically compute $U(\phi)$ by summing over all energy bands $\varepsilon_i(\phi,\bm{k}_\parallel)$ as
\begin{align}
U(\phi)=\sum_i\int d^2\bm{k}_\parallel\frac{\varepsilon_i(\phi,\bm{k}_\parallel)}{\exp[\varepsilon_i(\phi,\bm{k}_\parallel)/k_BT]-1}, \label{eq:U}
\end{align}
where $k_B$ is the Boltzmann constant and $T$ is the system temperature. Figure~\ref{fig:layers} plots $U(\phi)$ for the simplest possible situation---AF with only two atomic layers ($L=2)$---at $T=0.6$ and $\xi=0.6$. After a straightforward fitting, we see the following: (a) In the bilayer case, the $\mathcal{K}_1\cos2\phi$ term with $\mathcal{K}_1\approx24.17\times10^{-3}$ dominates, whereas all $\mathcal{K}_n$ components with $n>2$ are negligibly small; hence $\bm{S}_1$ prefers to be perpendicular to the N\'{e}el vector. (b) In the trilayer case, the leading contributions are $J_1\cos\phi+J_2\cos2\phi$ with $J_1=8.82\times10^{-3}$ and $\tilde{J}_2=J_2+\mathcal{K}_1=-1.58\times10^{-3}$. All $J_n$ with $n>3$ are negligibly small. The positive $J_1$ is a ferromagnetic exchange interaction between $\bm{S}_1$ and $\bm{S}_2$, whereas $|\tilde{J}_2|\approx 6.5\%\mathcal{K}_1\approx17.9\%J_1$ is a higher-order interlayer coupling; they both have counterparts in the electron-mediated interactions~\cite{ref:Grunberg,ref:Parkin,ref:Edwards,ref:Bruno,ref:tunnelRKKY1,ref:tunnelRKKY2}. 

\begin{figure*}[t]
	\centering
	\includegraphics[width=\linewidth]{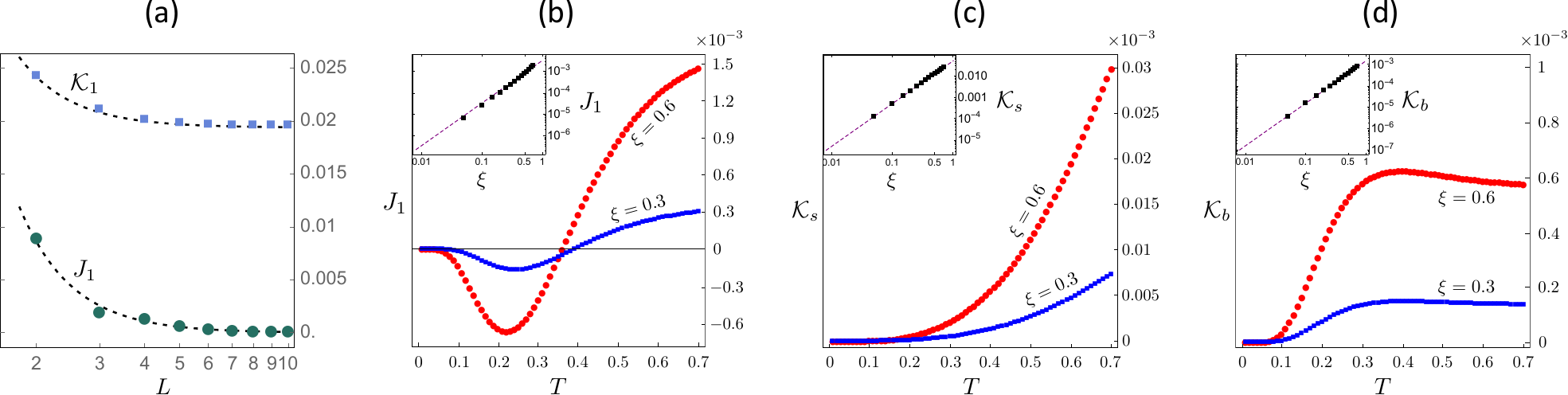}
	\caption{Properties of $J_1$ and $\mathcal{K}_1$. (a) The decay of $J_1$ fits into $1/L^3$; $\mathcal{K}_1$ fits into $\mathcal{K}_s+\mathcal{K}_b(L)$ with $\mathcal{K}_b\sim1/L^3$ and $\mathcal{K}_s$ being the interfacial contribution independent of $L$ (for $T=0.6$ and $\xi=0.6$). (b)--(d) plot the temperature dependences of $J_1$, $\mathcal{K}_s$ and $\mathcal{K}_b$, respectively (for $L=4$ and $\xi=0.3$ and $0.6$). The insets show the logarithmic plots of the $\xi$ dependences of $J_1$, $\mathcal{K}_s$ and $\mathcal{K}_b$ at $T=0.6$, which scales as $\xi^2$ (dashed lines).}
	\label{fig:data}
\end{figure*}

Next, we allow the AF to have multidomains where the N\'{e}el vector has a random in-plane orientation distributed uniformly from $0$ to $2\pi$. As a result, the magnon energy density of a given domain is $u=u(\phi,\gamma)$, where $\gamma$ specifies the local N\'{e}el order direction with respect to the reference [\textit{e.g.}, $\bm{S}_2$ in Fig.~\ref{fig:layers}(b)]. The total magnon energy becomes $U(\phi)=\int_0^{2\pi}\mathrm{d}\gamma u(\phi,\gamma)/2\pi$. Repeating the calculation of Eq.~\eqref{eq:U} for each individual domain and taking the multidomain average, we find the following: (a) In the bilayer case, the free F ($\bm{S}_1$) favors the out-of-plane direction since every domain in the AF prefers to align vertically with $\bm{S}_1$. The overall effect becomes a magnon-induced perpendicular anisotropy $\mathcal{K}_{\perp}=\mathcal{K}_1$. (b) In the trilayer case, the $J_2$ component drops out while $J_1$ survives, as the averaging operation introduces the azimuthal symmetry. Correspondingly, the interlayer coupling $J_1\cos\phi$ can be written as $J_1\bm{S}_1\cdot\bm{S}_2$. The above analysis indicates that $J_1$ and $K_1$ are the most relevant quantities in our geometry. In the following, we will focus on the physical properties of the two parameters.

When the AF becomes thicker, both $J_1$ and $K_1$ should get smaller because the fraction of spins under direct influence of boundaries reduces. Figure~\ref{fig:data}(a) plots $J_1$ and $\mathcal{K}_1$ versus the number of layers $L$ from $2$ to $10$. The decay patterns of $J_1$ and $\mathcal{K}_1$ fit well into
\begin{align}
&J_1\sim\frac{1}{L^3}, \\
&\mathcal{K}_1=\mathcal{K}_s+\mathcal{K}_b(L)\ \mbox{ with }\ \mathcal{K}_b(L)\sim\frac{1}{L^3}.
\end{align}
We interpret the $L$-independent term $\mathcal{K}_s$ as an interfacial anisotropy arising from surface magnons, which do not connect the two Fs together. By contrast, the term $\mathcal{K}_b(L)$ decays similarly as $J_1$; it originates from bulk magnons with (quantized) wave vectors in the thickness direction. We will discuss the physics underlying this $1/L^3$ decay of $J_1$ and $\mathcal{K}_b$ later.

Here we estimate the magnitudes of $J_1$ and $\mathcal{K}_1$. As shown earlier, $J_1$ is nearly $1\%$ of $J_{\rm AF}$ at $L=2$ assuming an interfacial coupling of $\xi=0.6$. With the $1/L^3$ decay, $J_1$ is roughly $0.1\%$ of $J_{\rm AF}$ at $L=4$. In insulating AFs such as Cr$_2$O$_3$~\cite{ref:Cr2O3}, $J_{\rm AF}$ is typically few tens of meV, and the lattice constant under the in-plane geometry is around $5$ $\mathrm{\AA}$. We then deduce that $J_1$ is $10-100$ $\mu$eV and $\mathcal{K}_1$ is $0.1-1$ meV at nanometer separations. Comparing with other interlayer couplings such as the Ruderman-Kittel-Kasuya-Yosida (RKKY) interaction~\cite{ref:Grunberg,ref:Parkin,ref:Edwards,ref:Bruno} and the magnetic dipole-dipole interaction~\cite{ref:Xiaodong} at similar separations, the magnon-mediated mechanism appears to be comparably strong.

Because magnons are bosons without particle number conservation (chemical potential is zero), they become more populated as we raise the temperature. As such, we would na\"{i}vely expect that the magnon-mediated interlayer coupling increases monotonically with temperature. In Figs.~\ref{fig:data}(b)--(d), we plot $J_1$, $\mathcal{K}_s$, and $\mathcal{K}_b$ as functions of temperature at separation $L=4$ for $\xi=0.3$ and $0.6$, where we see that only $\mathcal{K}_s$ bears a monotonic temperature dependence. Counterintuitively, $J_1$ flips sign at some finite temperature: it turns out to be antiferromagnetic at low temperatures and ferromagnetic at high temperatures. We find that the turning point shifts toward lower temperature for larger $\xi$ and thicker films (not shown), but this remarkable sign change persists. The underlying physics of this sign change will become clear later.

Considering that the electron-mediated RKKY interaction scales quadratically with the interfacial exchange coupling, we expect a similar behavior for the magnon-mediated interlayer coupling. The insets of Fig.~\ref{fig:data} show the $\xi^2$ fittings of $J_1$ and $\mathcal{K}_1$ in the range $\xi\in(0.05,0.65)$ with $0.05$ increment. Above $\xi=0.7$, the collinear ground state becomes energetically unstable such that the N\'{e}el vector undergoes a spin-flop transition and bends perpendicularly to the adjacent F, which invalidates our model. A stronger anisotropy $K_{\rm AF}$ can increase the spin-flop threshold, but it also enlarges the magnon gap of the AF. Since magnons are bosons that preferably occupy low-energy states, an enlarged gap will diminish the magnitude of our results (although the essential pattern in Fig.~\ref{fig:data} will be kept). It is numerically feasible to find a good balance between $K_{\rm AF}$ and the maximum possible $\xi$ without incurring spin-flop breakdown. Nevertheless, this is a matter of materials choice, and we leave it for future studies.

\begin{figure}[t]
	\includegraphics[width=0.98\columnwidth]{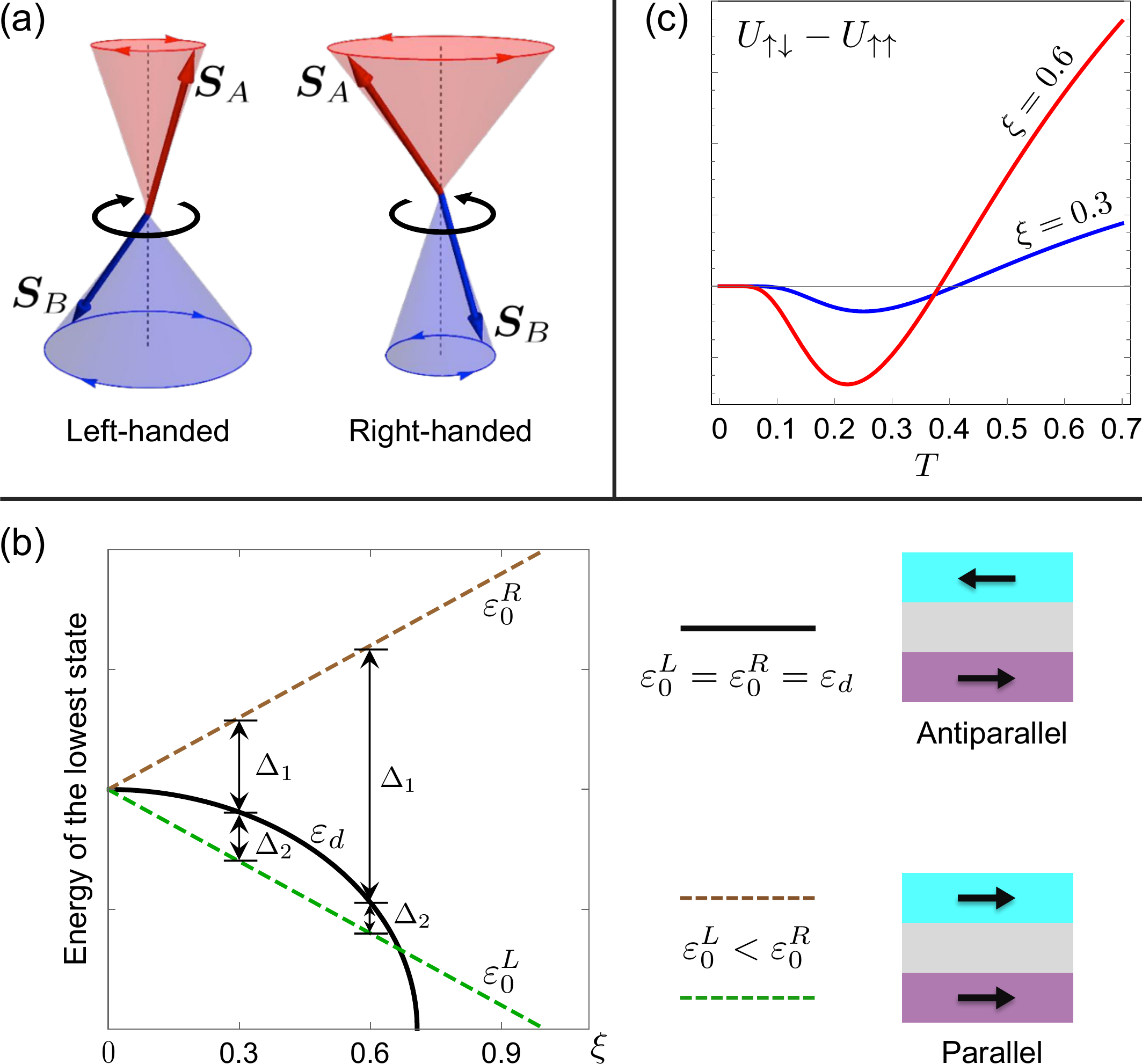}
	\caption{(a) Schematics of the two circularly polarized modes in the AF, where the two sublattice spins, $\bm{S}_A$ and $\bm{S}_B$, precess with the same chirality but different cone angles~\cite{ref:AFMR}. (b) Energy of the lowest magnon band as a function of the interfacial coupling $\xi$, where $\varepsilon_0^R$ ($\varepsilon_0^L$) stands for the right-handed (left-handed) subband. For parallel (antiparallel) alignment of the Fs, the degeneracy between $\varepsilon_0^R$ and $\varepsilon_0^L$ is lifted (maintained). For any $\xi$ below the spin-flop transition, the amplitude of energy splitting satisfies $\Delta_1>\Delta_2$. (c) $\Delta U=U_{\uparrow\downarrow}-U_{\uparrow\uparrow}$ as a function of $T$ for $\xi=0.3$ and $0.6$, which reproduces the essential feature of $J_1$ shown in Fig.~\ref{fig:data}(b).}
	\label{fig:explain}
\end{figure}

\textit{Discussions.}---The unique physical behavior of the magnon-mediated interlayer coupling is closely related to the Bose-Einstein statistics, where the most prominent contribution stems from a few low-energy states. Those states are long wavelength magnons ($\bm{k}_\parallel\rightarrow0$) with small band index $i$. Because of uniaxial symmetry, a magnon eigenstate is always circularly polarized and appears in pair with its partner of the opposite chirality~\cite{ref:Cr2O3,ref:AFMR}, as schematically depicted in Fig.~\ref{fig:explain}(a). At the bottom of the lowest band, all spins precess uniformly with nearly the same phase, and its left-handed (right-handed) component has energy $\varepsilon_0^L$ ($\varepsilon_0^R$). As shown in Fig.~\ref{fig:explain}(b), when the Fs are parallel and polarized along the easy axis of the AF, the degeneracy of the lowest band splits \textit{linearly} with $\xi$ (dashed lines). By contrast, when the Fs are antiparallel, the degeneracy is preserved so that $\varepsilon_0^L=\varepsilon_0^R=\varepsilon_d$ (solid black curve). As $\xi$ increases, $\varepsilon_d$ bends downward until reaching the spin-flop transition at around $\xi=0.7$. The energy relation between the parallel and antiparallel configurations is $\varepsilon_0^R=\varepsilon_d+\Delta_1$ and $\varepsilon_0^L=\varepsilon_d-\Delta_2$ with $\Delta_1>\Delta_2$. Accordingly, the difference of thermal energy between the two configurations reads
\begin{align}
 \Delta U=&U_{\uparrow\downarrow}-U_{\uparrow\uparrow} \notag\\
 =&\frac{2\varepsilon_d}{\exp(\varepsilon_d/k_BT)-1}-\frac{\varepsilon_d+\Delta_1}{\exp[(\varepsilon_d+\Delta_1)/k_BT]-1} \notag\\
    &\quad-\frac{\varepsilon_d-\Delta_2}{\exp[(\varepsilon_d-\Delta_2)/k_BT]-1}. \label{eq:diff}
\end{align}
Figure~\ref{fig:explain}(c) plots $\Delta U$ as a function of $T$ for $\xi=0.3$ and $0.6$, which qualitatively reproduces the numerical result in Fig.~\ref{fig:data}(b), suggesting that the non-monotonic behavior of $J_1$ is largely attributed to the lowest magnon band. Figure~\ref{fig:explain}(b) also reveals a fine structure: there is a very narrow window of $\xi$ just below the spin-flop transition within which $\varepsilon_d<\varepsilon_0^L<\varepsilon_0^R$; thus $\Delta_2$ becomes negative. In this special region, the antiparallel configuration always has a lower energy; thus the interlayer coupling is antiferromagnetic regardless of temperature.

Now we turn to the crucial question why $J_1$ scale as $1/L^3$. The magnon band structure along the thickness direction is analogous to that of a quantum well, where the gaps between neighboring energy levels scale as $1/L^2$. In addition, the percentage of boundary atoms that are exchange coupled to the Fs scales as $1/L$, \textit{i.e.}, the relative strength of exchange fields acting on the AF scales as $1/L$. Therefore, the magnon energy splitting ($\Delta_1$ and $\Delta_2$) scales as $1/L^3$. Up to the leading order in $\varepsilon/k_BT$, Eq.~\eqref{eq:diff} expands into $\Delta U=(\Delta_1-\Delta_2)/2+h.o.$, which respects the $1/L^3$ law. A similar argument applies to $\mathcal{K}_d$ as well; but in that case we should compare $U_{\uparrow\uparrow}$ with the thermal energy for the perpendicular F/AF configuration in which $\varepsilon_0^L=\varepsilon_0^R$ (no splitting). This provides a qualitative but not rigorous physical explanation of the decay pattern obtained numerically.

The conceived in-plane N\'{e}el vector shown in Fig.~\ref{fig:layers} has been realized in a recent experiment using Cr$_2$O$_3$~\cite{ref:Spinsuperfluid}. But it remains a challenge to achieve the collinear exchange-coupled F/AF interface based on this device geometry. Since our predicted effects scale quadratically in $\xi$, good interfacial quality is crucial. To fully verify our predictions, it calls for a serious materials search that goes beyond the scope of this Letter.

To close the discussion, we finally comment on the apparent distinction between the magnon-mediated interlayer coupling and its well-established electronic counterpart. Electrons obey the Fermi-Dirac statistics, hence most physical properties are directly linked to the existence of Fermi surface. Because of geometrical confinement in the thin film, electrons on the Fermi surface are subject to an oscillatory interference with an increasing spacer thickness, which gives rise to an oscillatory exchange coupling~\cite{ref:Edwards,ref:Bruno}. If the nonmagnetic spacer is electronically insulating while the two Fs are metallic, electron tunneling will become the dominant mechanism for the interlayer coupling~\cite{ref:Bruno,ref:tunnelRKKY1,ref:tunnelRKKY2}. In this case, the interlayer coupling does not oscillate but simply decays exponentially with the spacer thickness. In contrast to electrons, magnons are bosonic excitations that do not have a Fermi surface, which is responsible for the absence of oscillatory decay with thickness. The most defining feature, however, lies in the nontrivial sign change of $J_1$ with temperature, which can be directly measured from the hysteresis loop. Last but not least, we notice that the magnon-mediated interlayer coupling proposed in this Letter can be regarded as the magnonic counterpart of the thermal Casimir effect~\cite{ref:Casimir}.

In summary, we have demonstrated that two ferromagnetic layers separated by a thin-film AF can couple each other via magnon excitations inside the AF, which is especially important when all layers are insulating and conduction electrons are absent. The central result involves a magnon-mediated exchange interaction (a trilayer effect) and a magnon-induced anisotropy (a bilayer effect). Although the former decays with the AF thickness $L$ as $1/L^3$, the latter can be decomposed into a constant surface contribution independent of $L$ plus a term decaying as $1/L^3$. The interlayer exchange interaction turns out to be antiferromagnetic at low temperatures and ferromagnetic at high temperatures. This sign change can be qualitatively explained in terms of the lowest-band magnon excitation quantized in the thickness direction. Although antiferromagnetic magnons with opposite spins bear an apparent resemblance to electrons, the difference in their statistical properties leads to drastically different behaviors.

\begin{acknowledgments}
  R. C. acknowledges inspiring discussions with J. Shi (and his group members),  J. M. D. Coey, and D. Hou. R.C.\ and D.X.\ are supported by the Department of Energy, Basic Energy Sciences, Grant No.~DE-SC0012509.
\end{acknowledgments}


\begin{thebibliography}{20}
\bibitem{noteratio} The anisotropy is chosen to guarantee that the rotation of $\bm{S}_1$ does not drag the N\'{e}el order into a spiral so that the collinear ground state is not affected.
\end{thebibliography}
\end{document}